\documentclass[twocolumn,showpacs,prl]{revtex4}

\usepackage{graphicx}  
\usepackage{dcolumn}
\usepackage{amsmath}
\usepackage{amssymb}%
\begin{document}

\preprint{}

\title{Fluctuating particle motion during shear induced granular compaction}
 \author{O. Pouliquen}
\author{M. Belzons}
 \author{M. Nicolas}
\affiliation{Institut Universitaire des Syst\`{e}mes Thermiques et 
Industriels (IUSTI), 5 rue Enrico Fermi, 13453 Marseille cedex 13, France}

\date{\today}

\begin{abstract}
 Using a refractive index matching method, we investigate the trajectories of particles in a three dimensional granular packing submitted to cyclic shear deformation. The particle 
	motion observed during compaction is not diffusive but exhibits transient cage 
	effect, similar to the one observed in colloidal glasses. We precisely study the statistics of the step size between two successive cycles  and observe that  it is proportional to the shear amplitude. The link between the microscopic observations and the macroscopic evolution of the volume fraction during compaction is discussed. 
\end{abstract}
\maketitle
\vspace{1.5cm}

Granular material  are often described as athermal systems \cite{liu98,danna01}. For grains greater than few microns, thermal fluctuations are negligible  and play no role in the dynamics. A granular assembly can then be trapped in a metastable configuration and will not evolve unless an external perturbation such as vibration or shear is applied. The compaction of an initially loose granular sample  represents a simple configuration to study the dynamics of athermal system. 

Two configurations have been experimentally investigated to study  granular  compaction. The first one consists in imposing vertical taps to a granular packing \cite{Knight95,nowak98,josserand00,philippe03}. The second consists in imposing a periodic shear deformation to a sample contained in a box \cite{nicolas00,nicolas01}. In both cases, when the amplitude of excitation is constant (the tap intensity or the shear amplitude),  a slow logarithmic relaxation towards  more compact states is  observed \cite{Knight95,philippe03,nicolas00}.  In contrast to this slow dynamics,  rapid changes in volume fraction can be observed when a sudden change in excitation  is imposed \cite{josserand00,nicolas00}.  The packing dilates (resp. compacts) when the excitation amplitude increases (resp. decreases). Moreover the system exhibits a short term memory, i.e. two packings at the same volume fraction but being prepared differently do not evolve the same way when submitted to the same excitation. The coexistence of a slow and a rapid dynamics and the observation of memory effects have motivated number of numerical or theoretical models mainly based on free volume concept and geometrical constraints \cite{coniglio96,caglioti97,barker92,barrat01,boutreux97,head99}. However, to our knowledge, no link has been made between the macroscopic dynamics and the microscopic structure of the packing during compaction. Does the slow and rapid dynamics correspond to different grains rearrangements? What does the memory effect mean for the particles position? In this study we address these questions by analyzing the individual motion of the grains during a cyclic shear compaction.

In thermal system like colloidal suspensions, the microscopic  motion of the particles has been recently studied in detail. 
 Using confocal microscopy, Weeks {\it et al.}  \cite{weeks00,weeks02}  have been able to follow the  three dimensional trajectories of the particles in  colloidal hard spheres suspensions. They have shown that for dense systems near the glass transition the particles spend most of the time trapped in localized regions or "cages" and occasionally exhibit longer excursions. Considering the analogy often proposed between granular compaction and aging in glassy systems \cite{danna01}, one can legitimately wonder if similar microscopic behaviors are present in granular compaction.

 \begin{figure}[!ht]
  \begin{center}
  \includegraphics[scale=0.2]{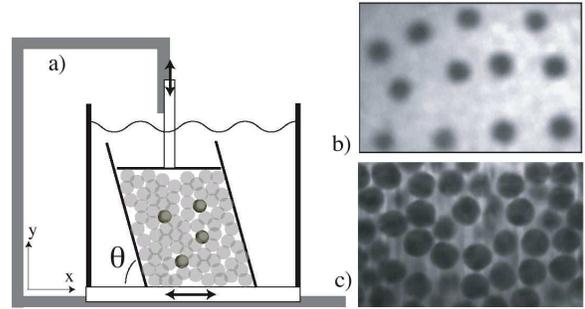}
  \caption{ a) Experimental setup. b)  Picture of the packing full of  index matching liquid with 12 colored tracers. The fuzziness allows a better accuracy for particle tracking. c) Picture obtained with the laser sheet method.}
  \end{center}
 \end{figure}

The experimental setup is described in details  in \cite{nicolas00} and is presented in Fig. 1. Spherical glass beads ($d=3$ mm in diameter)  contained in a  parallelepipedic box (7.7 cm wide, 10.2 cm deep and typically 10.5 cm 
high) are submitted to a periodic shear by inclining the two side walls. The top plate which applies a constant pressure on the packing is free to move vertically only. During compaction the top plate slowly goes down. The measure of its vertical position gives the bulk volume fraction. All the experiments are carried out in a quasistatic regime and  the only control parameter is the amplitude of the cyclic shear measured by the maximum inclination $\theta$ of the side walls.  

 \begin{figure}[!ht]
  \begin{center}
 \includegraphics[scale=0.3]{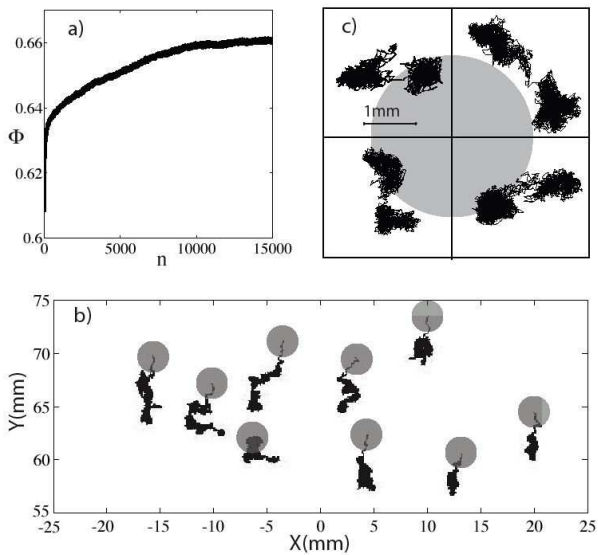}
  \caption{Compaction for $\theta=5.4^o$. a) Volume fraction as function of the number of cycles. b) Examples of trajectories during 15000 steps. The disks give the beads size and indicate the initial position of the tracers. c) Examples of cages (trajectories ploted for time slots between 2500 and 5000 steps). }
  \end{center}
 \end{figure}

The particles are tracked  during compaction using an index matching method. The material is
immersed in a mixture of  $67\%$ of turpentine and $33\%$ of methylnaphtalene  matching the glass refractive index of the beads. The packing is then transparent and it is possible to follow few colored glass beads initially introduced in the packing (Fig. 1b).  The experimental procedure is the following. The top plate being removed, the particles are gently poured in the container filled with the index matching liquid. Colored particles are then carefully introduced in the packing and the top plate is slowly put into contact with the beads. The mean initial volume fraction is typically 0.61. Once the packing is ready, the shear deformation is imposed:  the side walls are successively inclined to an angle $+\theta$ then $-\theta$, then  put back at the vertical position. After each cycle, we measure the volume fraction from the position of the top plate and we take a front picture of the packing. The tracers locations in the x-y plan are determined from the picture using a particle tracking algorithm. The location accuracy  is $\pm 15 \mu m$.  

In order to observe trajectory correlations between adjacent particles, another visualization technics is developed. Fluorescent dye is added to the mixture (with some alcohol to help dilution) and the packing is lighted with a vertical thin laser sheet in the x-y plane. One then obtains a picture where the particles appear as black disks (Fig 1c). This technics allows to study short time correlation motion but does not allow to follow particles in the long time, as they can move perpendicular to the laser plane and eventually leave it. 

The first experiments are performed at a constant shear amplitude $\theta$. An example of the  particle motion obtained for $\theta=5.4^o$ is presented in Fig. 2b. The plot represents the stroboscopic trajectory i.e. the successive positions of the particles measured after each shear cycle. 
Trajectories are first 
 characterized  by an overall decrease of the vertical coordinate as the number of cycles 
increases, reminiscent of the macroscopic compaction. The corresponding evolution of the volume fraction is plotted in Fig 2a. On top of this mean vertical displacement, one observes fluctuating motion characterized by ball-like regions as shown in the close-up Fig 2c. This could reveal 
 caging process. 
 The random motion of the particles seems to be trapped in a finite volume  before escaping and being trapped again in another cage. Notice that the change of cage  corresponds to a displacement  quite smaller than the particle diameter (about one fourth). When compaction goes on, the time spent in a cage becomes longer and longer and the particles eventually move around a fixed position.  The cage effect is reminiscent of what has been observed in colloidal glasses by Weeks {\it et al.} \cite{weeks00,weeks02}. The main difference here is that the system is not time invariant and evolves towards more and more compact states. Cage properties changing with time and  particle experiencing only few cage changes before being trapped in its final location (between 2 and 5), we have not been able to get significant statistics about the cage sizes. 

  \begin{figure}[!ht]
  \begin{center}
  \includegraphics[scale=0.27]{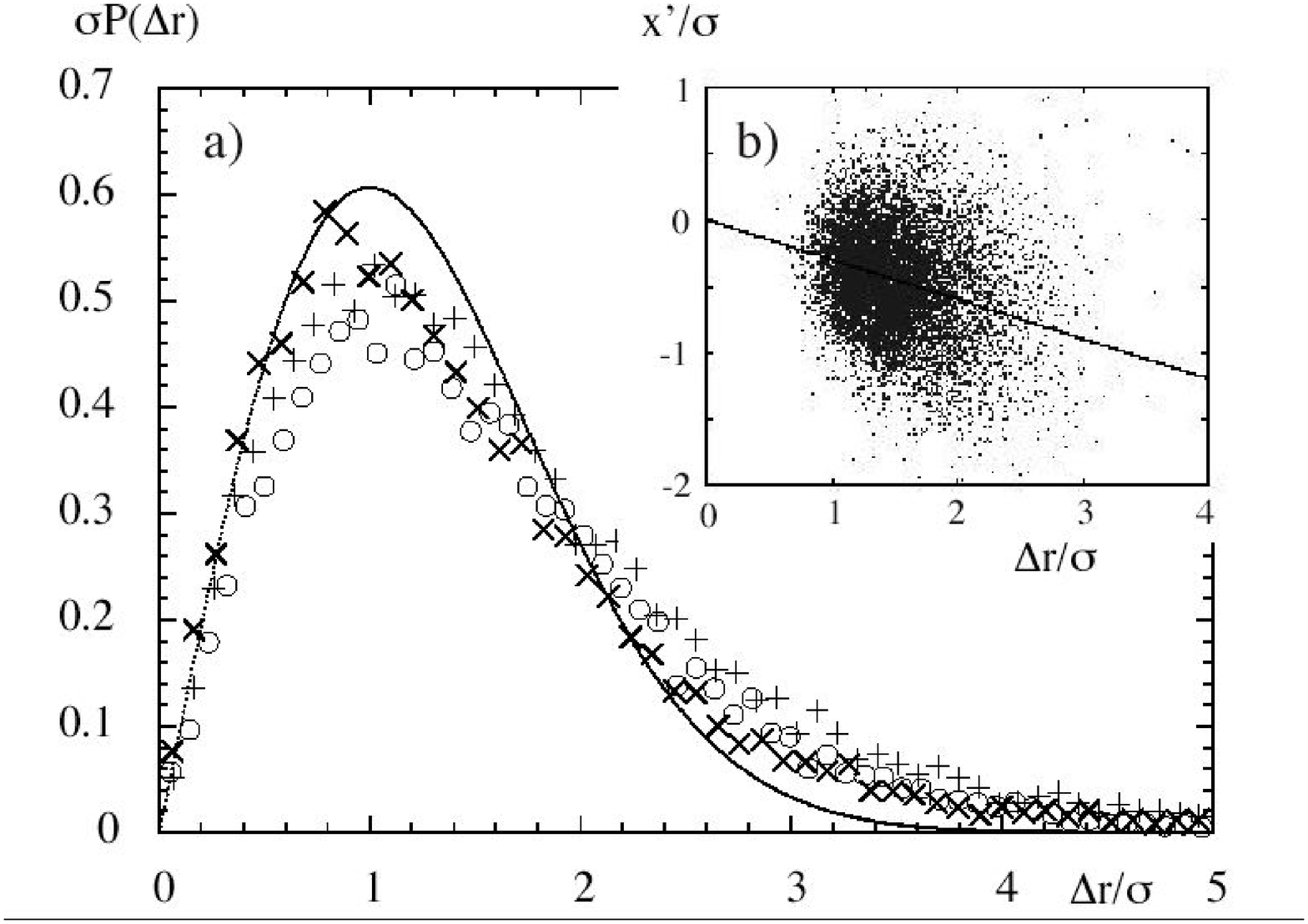}
  \caption{\label{fig:epsart} a) Normalized distribution  $P(\Delta r)$ computed at different stages during compaction; $\sigma$ is the standard deviation. $\theta=8.1^o$;
    ($\circ$) distribution obtained for $1<n<1000$ ($\sigma=72\mu m$); 
   (+) $4000<n<5000$  ($\sigma=54\mu m$);  ($\times$) $9000<n<10000$  ($\sigma=52\mu m$). Solid line is the gaussian 
  distribution. b) $x'$ as a function of  $\Delta r$ (see text). The line is a linear fit.} 
 \end{center}
 \end{figure}

However, in order to investigate how the random motion evolves with compaction, we study the statistics of the displacement $\Delta r=\sqrt{\Delta x^{2}+\Delta y^{2}}$ experienced by the particles between two successive cycles. The distribution of $P(\Delta r)$ obtained at various stages of the compaction process are plotted in Fig. 3, the step size being normalized by  the standard deviation $\sigma$. One observes that the distribution is slightly larger than that obtained if the $x$ and $y$ displacements were distributed according to a  gaussian centered on zero (solid line in Fig. 3). This observation is again very similar to the step size statistics observed in colloidal glasses \cite{weeks02}.

 Another analogy with colloidal glasses is observed when studying the correlation between two successive cycles : the random motion seems to be anti correlated.  When a particle moves in one direction, it has a great probability to move in the opposite direction the next step, which is reminiscent of a cage effect. This is shown in Fig. 3b where  the component $x' $ of the $n+1^{th}$  displacement along the direction of the $n^{th}$ displacement  is plotted against the size $\Delta r$ of  the $n^{th}$ displacement.  As in thermal colloidal glasses, $x'$ is statistically negative, revealing anti correlation. 

The granular compaction system  being not invariant in time,   we study how the step size distribution varies during compaction.  The different curves in Fig. 3 correspond to the distribution at  different time in an experiment at $\theta=8.1^o$ and show that  the shape of the normalized distribution does not vary much with time. However, the mean value $<\Delta r>$ varies  during the compaction as shown in Fig. 4.  The mean step size  $<\Delta r> $ is plotted as  a function of the volume fraction $\Phi$ during experiments carried out for three  different shear angles $\theta$.  In the three runs, the mean step displacement  first decreases down to $\Phi=0.63$ which corresponds roughly to 500 cycles and eventually reaches a constant value for dense enough packing.  The asymptotic value depends  on the shear amplitude and is smaller for small $\theta$. An interesting observation is that  once the packing is dense enough ($\Phi>0.64$), the  mean step displacement is only a function of $\theta$ independent of the history of the packing. This means that the same $<\Delta r>$ is measured if the compaction is carried out at a constant $\theta$ or if the packing is first compacted under a shear amplitude $\theta_1$ and then sheared at $\theta$. This is shown in Fig. 5  where we plot $<\Delta r> $ as a function of $\theta$ for many different experiments carried out at constant or non constant shear amplitude. The striking feature is the linear relation $<\Delta r> =\beta \theta$ with $\beta=8\mu m/deg$. The higher the shear amplitude, the larger the random displacement between two cycles. 

  \begin{figure}[!ht]
  \begin{center}
  \includegraphics[scale=0.27]{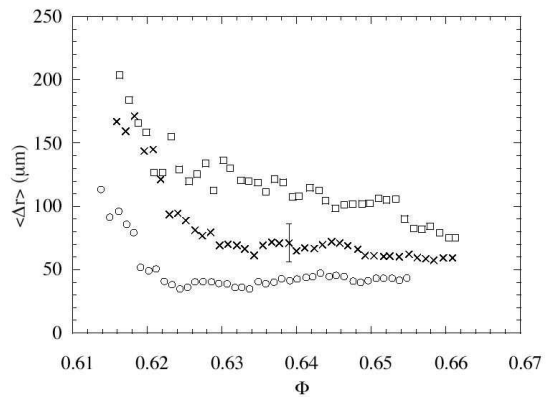}
  \caption{\label{fig:epsart} Evolution of the mean step size $<\Delta r>$ with $\Phi$ during compaction for $\theta=2.7^o (\circ) $; $ \theta=5.4^o (\times)$; $ \theta=8.4^o (\square)$. The average $<>$ means over 40 identical slots  in $\Phi$ and  over particles.  Fluctuations  from one particle to another due to heterogeneities in the packing and force network are given by the error bar. }
  \end{center}
 \end{figure}

This linear relation between microscopic displacement and  shear amplitude is reminiscent of the macroscopic observations of our previous study \cite{nicolas00}. When a sudden change $\Delta \theta$ was imposed to the packing, a rapid (in less than 10 cycles) variation of volume fraction $\Delta \Phi$ was measured. $\Delta \Phi$ was found to vary linearly with $\Delta \theta$ independently of the packing history for compacted enough packing. One can wonder if  a link exists between the microscopic random motion and the macroscopic behavior. 

\begin{figure}[!ht]
  \begin{center}
 \includegraphics[scale=0.27]{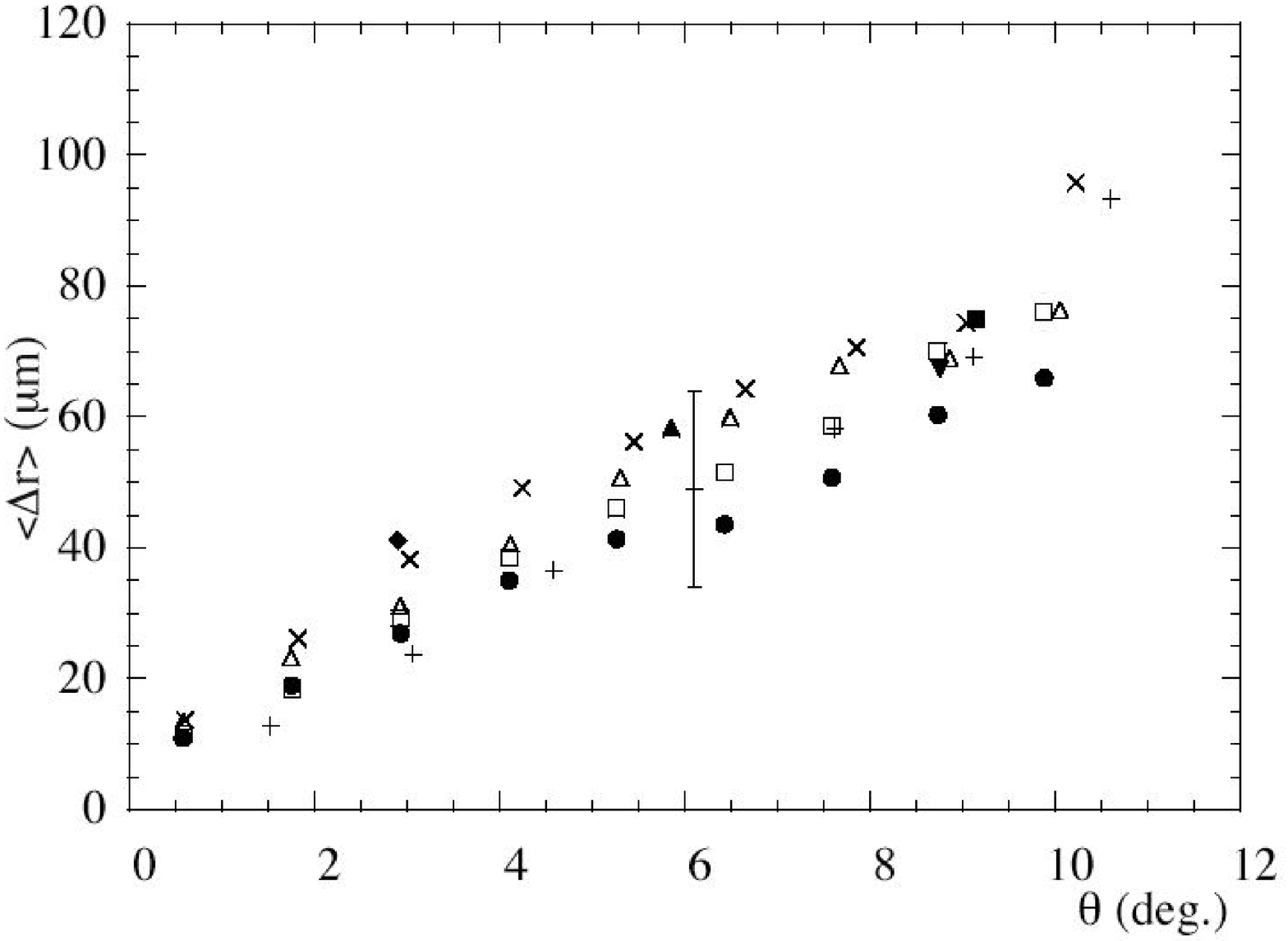}
  \caption{\label{fig:epsart} Mean step size $<\Delta r>$ as a function of shear angle $\theta$. The different symbols correspond to different experiments  with $\theta$ kept constant or changed  stepwise. }
  \end{center}
 \end{figure}

\begin{figure}[!ht]
  \begin{center}
  \includegraphics[scale=0.4]{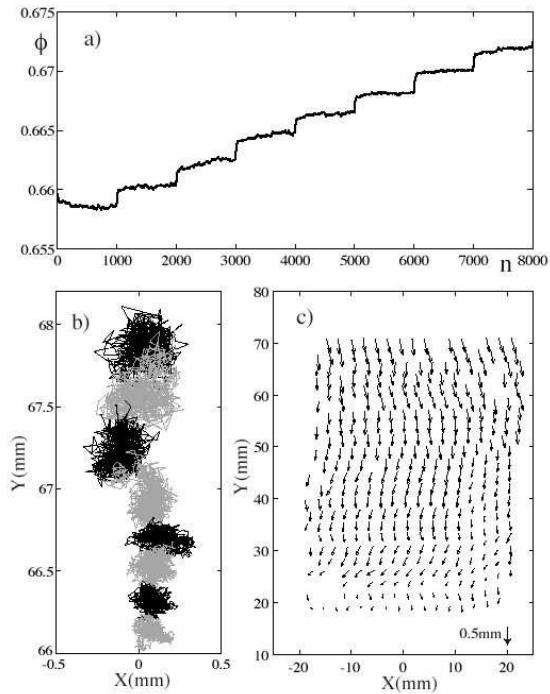}
  \caption{\label{fig:epsart} a) Volume fraction as a function of cycles when $\theta$ varies stepwise (see text).  b) Corresponding trajectory of one particle. Changes in color correspond to changes in $\theta$. c) Displacement field measured in the  cell when $\theta$ changes from 10.4$^o$ to 1.4$^o$. }
  \end{center}
 \end{figure}

In order to investigate this link, experiments are performed where the shear amplitude is discontinuously varied. A typical run is presented in Fig. 6. The packing is  first compacted at a fixed amplitude $\theta=8.4^o$ for $10^4$ cycles and then $\theta$ is varied from $10.2^o$ to $0.6^o$ by step of $1.2^o$, the angle being kept constant during 1000 cycles. The corresponding volume fraction variation is plotted in Fig 6a. As expected from our previous study \cite{nicolas00},  successive increasing steps in volume fraction are observed. The typical microscopic behavior of a particle during this experiment is presented in Fig 6b. The volume explored by the particle during its random motion successively shrinks when  the shear amplitude decreases because the mean step size $<\Delta r>$ decreases with $\theta$ (Fig. 5).  
However, each time the shear angle changes, the other particles below the test particle experience the same decrease in their exploration volume. The result is  a net downward motion observed when the angle changes. The observed volume fraction variation thus seems to result from the change in the volume randomly explored by  the particles. This becomes clear when looking at the displacement of all the particles during a change in  $\theta$  (Fig. 6c). Thanks to the laser light visualization, we are able to follow each particle in a plane when the shear angle changes from 10.4$^o$ to 1.4$^o$.  An homogeneous deformation is observed  (Fig. 6c), as if when the angle decreases, all the particles deflate yielding a change in volume fraction. An estimation of the change in volume fraction can be made. Assuming that the random fluctuating motion of a particle occupies a volume of the order $\left(d+<\Delta r> \right)^3$, the volume fraction is $\Phi =\Phi_0\frac{d^3}{\left(d+<\Delta r> (\theta)\right)^3}$ where $\Phi_0$ is a reference volume fraction. Knowing from Fig. 5 that $<\Delta r> =\beta \theta$, a change $\Delta \theta$ in the shear amplitude induces a volume fraction variation $\Delta \Phi = -3 \Phi_0 \beta  \Delta \theta$. A linear variation is predicted as observed in \cite{nicolas00} and the predicted coefficient (taking $\Phi_0=0.7$) is  $-3 \Phi_0 \beta=-0.56 \; 10^{-3}$deg$^{-1}$ having the right order of magnitude compared to the value  $-1.5  \; 10^{-3}$ deg$^{-1}$ measured in \cite{nicolas00}. 

All the results presented up to now are obtained for monodispersed glass beads. In this case, the compaction occurs towards crystallization \cite{nicolas00}.  We have checked that the microscopic behavior  does not qualitatively change when using a bidispersed material (mixture of 2 mm and 3 mm beads) which do not crystallize. Cages are still observed and the sudden change of volume fraction when changing the shear amplitude \cite{nicolas01} is  correlated to a change of the mean random step size.  The crystal nature of the packing then does not qualitatively influence the dynamics of the system.

In conclusion  a simple scenario can be proposed for the compaction process. We have seen that the particles motion can be divided in two kinds :  random motion within confined volumes or cages, whose extent is directly proportional to the shear amplitude, and occasionally longer exploration corresponding to a change of cage. We think that the slow dynamics of compaction observed in experiment at constant amplitude may be attributed to the change in cages. This changes are irreversible and push the system towards more and more compact configurations. How cage deformations occur, what is the influence of one cage rearrangement on the next ones  are open questions.  The cage changes certainly involve complex cooperative processes associated to modifications in the contact and force network which have not been studied here. 
   However, we have shown that the rapid change of volume fraction observed when changing the amplitude is of different nature:  it is simply related to the change of the cage size, without important structural changes. This means that this variation of volume is reversible and can be recovered by coming back  to the previous amplitude of excitation. The existence of these two processes which affect differently the packing volume fraction, explains that memory effects can be observed. 

 We thank F. Ratouchniak for his technical assistance.


\begin{thebibliography}{99}

\bibitem
{liu98} A. J. Liu and S. R. Nagel, Nature {\bf396}, 21 (1998).

\bibitem
{danna01} G. D'Anna and G. Gremaud, Nature {\bf413}, 407 (2001).

\bibitem
{Knight95} J. B. Knight, C. G. Fandrich, C. N. Lau, H. M. Jaeger and S. 
Nagel, Phys. Rev. E {\bf 51},  3957 (1995).

\bibitem
{nowak98} E. R. Novak, J. B. Knight, E. Ben-Naim, H. M. Jaeger and S. R. 
Nagel, Phys. Rev. E {\bf 57}, 1971 (1998).
	
\bibitem{josserand00} C. Josserand, A. Tkatchenko, D. Mueth and H. Jaeger, 
Phys. Rev. Lett. {\bf85}, 3632 (2000).

\bibitem
{philippe03} P. Philippe and D. Bideau,  Europhys. Lett. {\bf 60}, 
677 (2002).

\bibitem
{nicolas00} M. Nicolas, P. Duru and O. Pouliquen, Eur. Phys. J. E {\bf3}, 309 (2000).

\bibitem
{nicolas01} M. Nicolas, P. Duru and O. Pouliquen,  In {\it Powder and grains 2001}, edited by Y. Kishino  ( A.A. Balkema, Tokyo, 2001), p. 17-19.


\bibitem
{coniglio96} A. Coniglio and H. J. Herrmann, Physica A {\bf 225}, 
1 (1996).

\bibitem
{caglioti97} E. Caglioti, V. Loreto, H. J. Hermann and M. Nicodemi, A. Coniglio, Phys. Rev. Lett. {\bf 79}, 
1575 (1997).

\bibitem
{barker92} G. C. Barker and A. Mehta, Phys. Rev. A {\bf 45}, 
3435 (1992).

\bibitem
{barrat01} A. Barrat and V. Loreto Europhys. Lett. {\bf 53},297 (2001).

\bibitem
{boutreux97} T. Boutreux and P. G. de Gennes, Physica A{\bf 
244}, 
4758 (1997).
\bibitem
{head99} D. A. Head, Phys. Rev. E {\bf60}, 5685 (1999).

\bibitem
{weeks00}E. R. Weeks, J. C. Crocker, A. C. Levitt, A Schofield and D. A. Weitz, Science {\bf287}, 627 (2000).

\bibitem
{weeks02} E. R. Weeks and D. A. Weitz, Phys. Rev. Lett. {\bf89}, 095704 (2002). 


\end{thebibliography}
\end{document}